\documentclass[pra,twocolumn,amsmath,amssymb,superscriptaddress]{revtex4-1}
\usepackage{epsfig,amsmath}
\usepackage{subfigure}
\usepackage{graphicx}
\usepackage{dcolumn}
\usepackage{stmaryrd}
\usepackage{mathrsfs}
\usepackage{pifont}
\usepackage{amsthm}
\usepackage{amssymb}
\usepackage{bm}
\usepackage{latexsym}
\usepackage[colorlinks=true,linkcolor=blue,citecolor=blue]{hyperref}
\usepackage{color}
\usepackage{epstopdf}

\usepackage{booktabs}
\usepackage{threeparttable}
\usepackage{multirow}

\allowdisplaybreaks[3]

\begin{document}

\title{Universal quantum control with dynamical correction}

\author{Zhu-yao Jin}
\affiliation{School of Physics, Zhejiang University, Hangzhou 310027, Zhejiang, China}

\author{Jun Jing}
\email{Contact author: jingjun@zju.edu.cn}
\affiliation{School of Physics, Zhejiang University, Hangzhou 310027, Zhejiang, China}

\date{\today}

\begin{abstract}
Error correction is generally demanded in large-scale quantum information processing and quantum computation. We provide here a universal and realtime control strategy to dynamically correct the arbitrary type of errors in the system Hamiltonian. It yields multiple error-resilient paths for the interested system which are activated by the von Neumann equation for ancillary projection operators. With no extra control fields and precise designs, the path-dependent global phase alone suffices to mitigate the error-induced transitions among distinct paths as long as it varies faster than the other parameters. The corrected paths can also be regarded as the approximate solutions to the time-dependent Schr\"odinger equation perturbed by errors. Our dynamical-correction strategy is practiced with the cyclic transfer of populations in a three-level system, showing a superior error resilience to the parallel-transport condition. It provides a promising idea for advancing control methodologies in imprecise quantum systems.
\end{abstract}

\maketitle

\section{Introduction}

Quantum computing has advantages over its classical counterpart in solving certain problems~\cite{John2018Quantum,Sergey2018Quantum}, e.g., Shor's algorithm for factoring large numbers~\cite{Ekert1996Quantum,Kivlichan2020Improved}. However, factoring a $2048$-bit Rivest-Shamir-Adleman (RSA) integer demands $20$ million physical qubits and $2.7$ billion Toffoli gate with error rates of $10^{-10}$ per gate~\cite{Gidney2019HowTF}. In sharp contrast, state-of-the-art quantum processors typically exhibit error rates of $10^{-3}$ per gate~\cite{Gaebler2016High,Rol2019Fast,Foxen2020Demonstrating,Wu2021Strong}, which remain far too high for execution of a practical circuit. Without a fault-tolerant design, quantum computation would be restricted to problems so small that they would be of trivial use or interest. With quantum error correction (QEC)~\cite{Shor1995Scheme}, quantum computation is strongly believed to be scaled up enough to allow powerful next-generation calculations. The key to QEC is to identify the error resources and then to compensate for their effect.

In the first QEC scheme~\cite{Shor1995Scheme}, nine physical qubits were used to encode one logical qubit to address the errors caused by environmental decoherence. Due to the redundancy of the large Hilbert space, the failure of any physical qubit does not corrupt the underlying logical information and it can be unambiguously detected and corrected in real time~\cite{Nielsen2010Quantum}. Inspired by Shor's code, many QEC coding schemes~\cite{Egan2021Fault,Ryan2021Realization,Abobeih2021Fault,Sundaresan2023Demonstrating,Krinner2022Realizing,
Zhao2022Realization,Ofek2016Extending,Fluhmann2019Encoding,Campagne2020Quantum,Grimn2020Stabilization} have been proposed, including the Bacon-Shor~\cite{Egan2021Fault}, color~\cite{Ryan2021Realization}, five-qubit~\cite{Abobeih2021Fault}, heavy-hexagon~\cite{Sundaresan2023Demonstrating}, surface~\cite{Krinner2022Realizing,Zhao2022Realization}, and continuous-variable~\cite{Ofek2016Extending,Fluhmann2019Encoding,Campagne2020Quantum,Grimn2020Stabilization} codes. Their usefulness can be assessed using the so-called break-even point~\cite{Google2023Suppressing,Ni2023Beating,Paetznick2024Demonstration}. In other words, the lifetime of the encoded logical qubits should be larger than that of the best available physical qubits. The excess over the break-even point was first demonstrated in the superconducting qubit system~\cite{Google2023Suppressing}, followed by the quantum electrodynamics architecture~\cite{Ni2023Beating} and the trapped-ion system~\cite{Paetznick2024Demonstration}.

Another major source of errors is the fluctuation in system parameters, which arises from imperfect knowledge of the system or generic variations in the experimental parameters during practical implementation. As an early approach in nuclear magnetic resonance~\cite{Levitt1986Composite}, the composite-pulse method~\cite{Tycko1983Broadband,Vitanov2011Arbitrarily,Wang2012Composite,Kyoseva2013Arbitrarily,Kestner2013Noise,
Wang2014Robust,Genov2014Correction,Yang2018Neural,Torosov2019robust,Kyoseva2019Detuning,Shi2021Robust,Wu2023Composite} adopted multiple precisely timed pulses to compensate for the systematic errors in the time duration~\cite{Wang2012Composite,Kestner2013Noise,Yang2018Neural}, the detuning~\cite{Torosov2019robust,Kyoseva2019Detuning}, and the phase~\cite{Vitanov2011Arbitrarily,Kyoseva2013Arbitrarily,Genov2014Correction}. To avoid the high overhead associated with multiple pulses, the single-shot pulse method~\cite{Ruschhaupt2012Optimally,Daems2013Robust,Liang2024Nonadiabatic} based on perturbation theory, was proposed to locate and compensate for systematic errors by adjusting the global phase. This method can be interpreted with geometry in Hilbert space~\cite{Dridi2020Optimal,Dridi2024Optimal,Propson2022Robust}. As a general dynamical-decoupling method, nonperturbative leakage elimination operators~\cite{Jing2015Nonperturbative} were introduced to dynamically leverage the errors with a sufficiently large integral of the pulse sequence in the time domain. Also, resistance to systematic errors can be strengthened by nonadiabatic controls, including the digital optimization method~\cite{Harutyunyan2023Digital}, the numerical optimization algorithm~\cite{Zhang2019Searching}, the dark-path schemes~\cite{Ming2022Experimental,Andr2022Dark,Jin2024Geometric}, the measurement-based mitigation method~\cite{Zhao2024Mitigation}, and the pulse-shaping method~\cite{Liu2019Plug}. Most of them focus on systems with a few levels or those with SU(2) dynamical symmetry~\cite{Genov2011Optimized} and their application range is restricted to certain type of errors. A universal control framework applicable to discrete systems of any dimension and resilient to arbitrary systematic error induced by control is apparently required.

In this paper, we develop a forward-engineering framework for dynamically correcting the arbitrary type of systematic errors. Inspired by the universal quantum control of the time-dependent Hamiltonian~\cite{Jin2024Shortcut}, the system dynamics is prescribed in a rotated picture spanned by ancillary basis states, which can be activated as nonadiabatic paths by the von Neumann equation for the ancillary projection operators. In the absence of systematic errors, no transition occurs among the nonadiabatic paths during the time evolution. In nonideal situations, the errors arising in the control parameters can be described in the second-rotated picture using the Magnus expansion. The global phase of a desired path is found to be a crucial degree of freedom to suppress unwanted transitions to other paths, irrespective of the error type. It has seldom been discussed in relation to conventional control methods, such as shortcuts to adiabaticity~\cite{Chen2011Lewis,Guery2019Shortcuts,Qi2022Accelerated,Chen2010Shortcut,Jing2016Eigenstate,Baksic2016Speeding,
Guery2019Shortcuts} and the nonadiabatic holonomic transformation~\cite{Sjoqvist2012Nonadiabatic,Liu2019Plug} under the parallel-transport condition~\cite{Jing2017Non,Zheng2016Comparison}. Our dynamical-correction theory is further illustrated by the cyclic population transfer in a ubiquitous three-level system.

The rest of this paper is structured as follows. In Sec.~\ref{Ideal}, we recall our universal control framework that generates multiple nonadiabatic paths under an error-free time-dependent Hamiltonian. In Sec.~\ref{Nonideal}, it is extended to the nonideal situation of a general erroneous Hamiltonian. Arbitrary type of systematic errors are identified within a second-rotated picture and then suppressed to the second order of error magnitude. In Sec.~\ref{IdealThree}, we provide a detailed construction of universal passages for a general three-level system in the absence of errors. In Sec.~\ref{NonidealThree}, we practice our dynamical-correction theory in the presence of either global or local errors by rapidly manipulating the path-dependent global phase. Section~\ref{PopErr} examines our correction mechanism during the cyclic transfer of populations of the three-level system subject to errors. The work is concluded in Sec.~\ref{conclusion}. Appendixes~\ref{DetailEvo} and \ref{DetailOverlap} provide detailed derivations of the perturbed time-evolution operator and the overlap between the perturbed state and the target state, respectively.

\section{General framework}\label{general}

In this section, the dynamical correction addressing systematic errors in the control parameters is performed under our universal framework~\cite{Jin2024Shortcut}, which was proposed to unify the transitionless dynamics~\cite{Berry2009Transition}, the shortcuts to adiabaticity~\cite{Chen2011Lewis,Guery2019Shortcuts,Qi2022Accelerated,Chen2010Shortcut,Baksic2016Speeding}, the holonomic quantum transformation~\cite{Sjoqvist2012Nonadiabatic,Liu2019Plug}, and the reverse-engineering method~\cite{Daems2013Robust,Golubev2014Control,Gonz2017Invariant,Dridi2020Optimal}. It was developed from the idea of a quantum generalization of the d'Alembert principle for virtual displacement. We first briefly review the general theory with an error-free Hamiltonian and then extend it to nonideal situations with systematic errors.

\subsection{Ideal situation}\label{Ideal}

The universal perspective on nonadiabatic quantum control~\cite{Jin2024Shortcut} was first discussed for a closed quantum system of arbitrary finite size driven by a well-defined and time-dependent Hamiltonian $H_0(t)$. The system dynamics is given by the time-dependent Schr\"odinger equation as ($\hbar\equiv1$)
\begin{equation}\label{Sch}
i\frac{d|\psi_m(t)\rangle}{dt}=H_0(t)|\psi_m(t)\rangle,
\end{equation}
where $|\psi_m(t)\rangle$'s are the pure-state solutions. Suppose that the Hilbert space of the system is $K$ dimensional; then $m$ runs from $1$ to $K$ when $H_0(t)$ is full rank. The time-evolution operator $U_0(t)$ with respect to $H_0(t)$ is generally hard to obtain unless $|\psi_m(t)\rangle$'s have already constituted an orthonormal set for the Hilbert space of the system.

Alternatively, the time-dependent Schr\"odinger equation~(\ref{Sch}) can be treated within a rotated picture spanned by the ancillary basis states $|\mu_k(t)\rangle$'s, with $1\leq k\leq K$, that span the same Hilbert space as $|\psi_m(t)\rangle$'s. In the rotating frame with respect to $V(t)\equiv\sum_{k=1}^K|\mu_k(t)\rangle\langle\mu_k(0)|$, Eq.~(\ref{Sch}) is transformed to be
\begin{equation}\label{Schrot}
i\frac{d|\psi_m(t)\rangle_{\rm rot}}{dt}=H_{\rm rot}(t)|\psi_{m}(t)\rangle_{\rm rot},
\end{equation}
where the rotated pure-state solutions $|\psi_{m}(t)\rangle_{\rm rot}$ are written as
\begin{equation}\label{relate}
|\psi_m(t)\rangle_{\rm rot}=V^\dagger(t)|\psi_m(t)\rangle
\end{equation}
and the rotated system Hamiltonian $H_{\rm rot}(t)$ reads
\begin{equation}\label{Hamrot}
\begin{aligned}
&H_{\rm rot}(t)=V^\dagger(t)H_0(t)V(t)-iV^\dagger(t)\frac{d}{dt}V(t)\\
&=-\sum_{k=1}^K\sum_{n=1}^K\left[\mathcal{G}_{kn}(t)-\mathcal{D}_{kn}(t)\right]|\mu_k(0)\rangle\langle\mu_n(0)|,
\end{aligned}
\end{equation}
where $\mathcal{G}_{kn}(t)\equiv i\langle\mu_k(t)|\dot{\mu}_n(t)\rangle$ and $\mathcal{D}_{kn}(t)\equiv\langle\mu_k(t)|H_0(t)|\mu_n(t)\rangle$ represent the geometrical and dynamical components of the matrix elements, respectively.

Certain conditions~\cite{Berry2009Transition,Vitanov2017Stiumlated} have to be imposed to analytically solve Eq.~(\ref{Schrot}). When the Hamiltonian $H_{\rm rot}(t)$ in Eq.~(\ref{Hamrot}) is diagonalized, i.e., $\mathcal{G}_{kn}(t)-\mathcal{D}_{kn}(t)=0$ for $k\ne n$, Eq.~(\ref{Hamrot}) is simplified to
\begin{equation}\label{Hdigfull}
H_{\rm rot}(t)=-\sum_{k=1}^K\left[\mathcal{G}_{kk}(t)-\mathcal{D}_{kk}(t)\right]|\mu_k(0)\rangle\langle\mu_k(0)|.
\end{equation}
Consequently, the time-evolution operator $U_{\rm rot}(t)$ can be directly obtained from Eq.~(\ref{Hdigfull}) as
\begin{equation}\label{Urot1}
U_{\rm rot}(t)=\sum_{k=1}^{K}e^{if_k(t)}|\mu_k(0)\rangle\langle\mu_k(0)|,
\end{equation}
where the global phases $f_k(t)$'s are defined as
\begin{equation}\label{global}
f_k(t)\equiv\int_0^t\left[\mathcal{G}_{kk}(t_1)-\mathcal{D}_{kk}(t_1)\right]dt_1.
\end{equation}
According to Eq.~(\ref{relate}), the time-evolution operator $U_0(t)$ in the original picture is found to be
\begin{equation}\label{U0}
U_0(t)=V(t)U_{\rm rot}(t)=\sum_{k=1}^{K}e^{if_k(t)}|\mu_k(t)\rangle\langle\mu_k(0)|.
\end{equation}
Equation~(\ref{U0}) implies that if the system tracks the starting point of the path $|\mu_k(0)\rangle$, it will stick to the instantaneous state $|\mu_k(t)\rangle$ and accumulate a global phase $f_k(t)$, with no transitions to the other paths $|\mu_{n\neq k}(t)\rangle$ during the time evolution.

It was proved that the diagonalization of $H_{\rm rot}(t)$ in Eq.~(\ref{Hdigfull}) is ensured by the von Neumann equation~\cite{Liu2019Plug,Jin2024Shortcut}
\begin{equation}\label{von}
\frac{d}{dt}\Pi_k(t)=-i\left[H_0(t), \Pi_k(t)\right],
\end{equation}
with $H_0(t)$ being the system Hamiltonian and the ancillary projection operator $\Pi_k(t)\equiv|\mu_k(t)\rangle\langle\mu_k(t)|$. The $k$th path $|\mu_k(t)\rangle$ is found to be useful in state control unless it becomes a time-independent dark state or dark mode, i.e., $|\mu_k(t)\rangle\rightarrow|\mu_k\rangle$ due to $H_0(t)|\mu_k\rangle=0$~\cite{Jin2024Entangling}. One can find a recipe in Ref.~\cite{Jin2024Entangling} to construct the basis states $|\mu_k(t)\rangle$'s for a general two-band system. Equation~(\ref{Anci}) presents the ansatz of $|\mu_k(t)\rangle$'s for an arbitrary three-level system.

\subsection{Nonideal situation}\label{Nonideal}

When the dynamics are under control, systematic error can occur in the Rabi frequencies~\cite{Wang2012Composite,Kestner2013Noise,Yang2018Neural}, the driving frequencies~\cite{Torosov2019robust,Kyoseva2019Detuning}, and the phases~\cite{Vitanov2011Arbitrarily,Kyoseva2013Arbitrarily,Genov2014Correction} of the driving fields, typically induced by the random variation of the experimental parameters. In general, the system evolves according to
\begin{equation}\label{SchNon}
i\frac{d|\psi_m(t)\rangle}{dt}=[H_0(t)+\epsilon H_1(t)]|\psi_m(t)\rangle,
\end{equation}
where $H_0(t)$ is the ideal Hamiltonian, $H_1(t)$ is the error Hamiltonian, and $\epsilon$ serves as an error magnitude.

Like for the ideal situation in Sec.~\ref{Ideal}, we start with the rotation with respect to $V(t)=\sum_{k=1}^K|\mu_k(t)\rangle\langle\mu_k(0)|$. Then the rotated Hamiltonian is rewritten as
\begin{equation}\label{HamrotDiag}
\begin{aligned}
H_{\rm rot}^{\rm err}(t)&=-\sum_{k=1}^K\dot{f}_k(t)|\mu_k(0)\rangle\langle\mu_k(0)|\\
&+\epsilon\sum_{k=1}^K\sum_{n=1}^K\mathcal{D}_{kn}^{\rm err}(t)|\mu_k(0)\rangle\langle\mu_n(0)|,
\end{aligned}
\end{equation}
where $\mathcal{D}_{kn}^{\rm err}(t)\equiv\langle\mu_k(t)|H_1(t)|\mu_n(t)\rangle$. The second term in Eq.~(\ref{HamrotDiag}) describes the unwanted transitions between different paths induced by the error Hamiltonian $H_1(t)$. They can be addressed in the second-rotated picture. With respect to the unitary rotation $U_{\rm rot}(t)$ in Eq.~(\ref{Urot1}), the system dynamics can be described by
\begin{equation}\label{Schrot2}
i\frac{d|\psi_m(t)\rangle_I}{dt}=H_I(t)|\psi_{m}(t)\rangle_I,
\end{equation}
where the second-rotated pure-state solution $|\psi_{m}(t)\rangle_I$ reads
\begin{equation}\label{relate2}
|\psi_m(t)\rangle_I=U_{\rm rot}^\dagger(t)|\psi_m(t)\rangle_{\rm rot}
\end{equation}
and the second-rotated system Hamiltonian $H_I(t)$ can be written as
\begin{equation}\label{HamI}
\begin{aligned}
H_I(t)&=U_{\rm rot}^\dagger(t)H_{\rm rot}^{\rm err}(t)U_{\rm rot}(t)-iU_{\rm rot}^\dagger(t)\frac{d}{dt}U_{\rm rot}(t)\\
&=\epsilon\sum_{k=1}^K\sum_{n=1}^K\tilde{\mathcal{D}}_{kn}^{\rm err}(t)|\mu_k(0)\rangle\langle\mu_n(0)|\equiv\epsilon\tilde{\mathcal{D}}(t)
\end{aligned}
\end{equation}
with
\begin{equation}\label{ErrElement}
\tilde{\mathcal{D}}_{kn}^{\rm err}(t)\equiv\langle\mu_k(t)|H_1(t)|\mu_n(t)\rangle e^{-i[f_k(t)-f_n(t)]}.
\end{equation}
Here $\tilde{\mathcal{D}}(t)$ is the error Hamiltonian expressed with the ancillary basis states $|\mu_k(0)\rangle$'s.

For the time-dependent Schr\"odinger equation~(\ref{Schrot2}), the time-evolution operator can be given by
\begin{equation}\label{Uformu}
U_I(t)=\hat{\mathcal{T}}e^{-i\int_0^tH_I(t')dt'},
\end{equation}
where $\hat{\mathcal{T}}$ is the time-order operator. For the Magnus expansion~\cite{Blanes2009Magnus,Liu2021Super} about  Eq.~(\ref{Uformu}), we have
\begin{equation}\label{Umagnus}
U_I(t)=\exp\left[\sum_{l=1}^{\infty}\Lambda_l(t)\right],
\end{equation}
where the first and second order elements are
\begin{equation}
\begin{aligned}
\Lambda_1(t)&=-i\int_0^tdt_1H_I(t_1),\\
\Lambda_2(t)&=\frac{(-i)^2}{2}\int_0^tdt_1\int_0^{t_1}dt_2\left[H_I(t_1), H_I(t_2)\right],
\end{aligned}
\end{equation}
respectively. Under the assumption that $\epsilon$ is a time-independent and perturbative coefficient~\cite{Vitanov2011Arbitrarily,Wang2012Composite}, i.e., $\epsilon\ll1$, the evolution operator in Eq.~(\ref{Umagnus}) can be expanded to the second order of $\epsilon$~\cite{Blanes2009Magnus} (see Appendix~\ref{DetailEvo} for details):
\begin{equation}\label{Uapprox}
\begin{aligned}
U_I(t)&\approx1-i\epsilon\mathcal{M}(t)\\
&-\frac{\epsilon^2}{2}\left\{\mathcal{M}(t)^2+\int_0^tdt_1\int_0^{t_1}dt_2
\left[\tilde{\mathcal{D}}(t_1), \tilde{\mathcal{D}}(t_2)\right]\right\},
\end{aligned}
\end{equation}
where $\mathcal{M}(t)$ is the time integral of the error Hamiltonian $\tilde{\mathcal{D}}(t)$ in the ancillary picture
\begin{equation}\label{Convension}
\mathcal{M}(t)\equiv\int_0^tdt_1\tilde{\mathcal{D}}(t_1),
\end{equation}
and it can be regarded as an error unitary rotation.

During the evolution along the path $|\mu_k(t)\rangle$, the impact of the systematic errors can be estimated by the overlap or fidelity between the instantaneous state under $H_I(t)$ and the target state (see Appendix~\ref{DetailOverlap} for details),
\begin{equation}\label{overlap}
\begin{aligned}
&\mathcal{F}\equiv\left|\langle\mu_k(0)|U_I(t)|\mu_k(0)\rangle\right|^2\\
&\approx\left|1-i\epsilon\mathcal{M}_{kk}(t)-\epsilon^2\sum_{n=1}^K\int_0^tdt_1\mathcal{M}_{nk}(t_1)
\tilde{\mathcal{D}}_{kn}^{\rm err}(t_1)\right|^2\\
&=1-\epsilon^2\sum_{n=1,n\ne k}^K\left|\mathcal{M}_{kn}(t)\right|^2+\mathcal{O}\left(\epsilon^3\right),
\end{aligned}
\end{equation}
where the square of the off-diagonal elements $|\mathcal{M}_{kn}(t)|^2$, with $k\ne n$, indicate the probability of the unwanted transition from the desired path $|\mu_k(t)\rangle$ to the other paths $|\mu_n(t)\rangle$'s. When $K=2$, Eq.~(\ref{overlap}) recovers the results obtained with the perturbation theory for two-level systems~\cite{Daems2013Robust,Dridi2020Optimal,Liang2024Nonadiabatic}. It is straightforward to see that the error can be suppressed to the second order of $\epsilon$ when
\begin{equation}\label{errorcondition}
\begin{aligned}
&\left|\mathcal{M}_{kn}(t)\right|^2\\
=&\left|\int_0^t\langle\mu_k(t_1)|H_1(t_1)|\mu_n(t_1)\rangle e^{-i[f_k(t_1)-f_n(t_1)]}dt_1\right|^2\approx0.
\end{aligned}
\end{equation}

\emph{A brief recipe for dynamical correction.} Equation~(\ref{errorcondition}) holds under at least three mechanisms: (1) The integral kernel of $\mathcal{M}_{kn}(t)$ can be divided into multiple and even infinite segments, and the integral over each segment can be nullified under the adiabatic condition~\cite{Vitanov2017Stiumlated}. (2) The kernel itself is a periodic function~\cite{Daems2013Robust,Dridi2020Optimal} or can be designed to be an inversely symmetrical function~\cite{Liu2021Super} such that the integral over the entire time domain vanishes. (3) By appropriately setting the global phases $f_k(t)$'s so that $\langle\mu_k(t)|H_1(t)|\mu_n(t)\rangle$ is a slowly varying function with time in comparison to the exponential function $\exp[-if_k(t)+if_n(t)]$, the overall integral is neutralized~\cite{Jing2015Nonperturbative}. In other words, if we have
\begin{equation}\label{OptGeneral}
\left|\dot{f}_k(t)-\dot{f_n}(t)\right|\gg\frac{d}{dt}\left[\langle\mu_k(t)|H_1(t)|\mu_n(t)\rangle\right],
\end{equation}
then the systematic errors can be corrected. To some extent, the dynamical-correction mechanism 3 relaxes the strict requirement over multiple vanishing segments of the integral in mechanism 1 and avoids the precise design of the periodic behavior exhibited by the integral kernel in mechanism 2.

The dynamical-correction condition in Eq.~(\ref{OptGeneral}) is mathematically supported by the Riemann-Lebesgue lemma, which states that if $\langle\mu_k(t)|H_1(t)|\mu_n(t)\rangle$ as a function of time is Lebesgue integrable within an interval, then its Fourier coefficient such as the integral in Eq.~(\ref{errorcondition}) approaches vanishing with a rapidly-varying phase difference $|\dot{f}_k(t)-\dot{f}_n(t)|$. The time-evolution operator in Eq.~(\ref{Uapprox}) is found to be close to the identity operator in the second rotated picture under the correction condition, irrespective of the formation of the error Hamiltonian $H_1(t)$, the magnitude of $\epsilon$, and even the number of nontrivial paths satisfying the von Neumann equation~(\ref{von}).

For a large-scale discrete-variable system, one can simply set the global phase $f_k(t)$ of the desired path $|\mu_k(t)\rangle$ as a fast-varying function of time and set all the other phases $f_{n\neq k}(t)$'s to be constant with time. Thus the system dynamics along the nonadiabatic path $|\mu_k(t)\rangle$ can feature a nearly unit fidelity, as indicated by Eqs.~(\ref{overlap}), (\ref{errorcondition}), and (\ref{OptGeneral}). We therefore have at least one near-perfect path and our universal dynamical correction is scale-free. If two paths $|\mu_{k_1}(t)\rangle$ and $|\mu_{k_2}(t)\rangle$ are required for a certain control task, then one can set $f_{k_1}(t)=-f_{k_2}(t)=f(t)$ and $f_n(t)=f_n$ for $n\neq k_1,k_2$, where $f(t)$ satisfies Eq.~(\ref{OptGeneral}). This highlights the broad applicability of our correction mechanism. In contrast, the correction mechanism (1) and (2) are hardly practical for a high-dimensional system and are applicable only to certain types of errors.

\section{An Illustrative example}\label{Optimize}

\subsection{Ideal situation}\label{IdealThree}

\begin{figure}[htbp]
\centering
\includegraphics[width=0.85\linewidth]{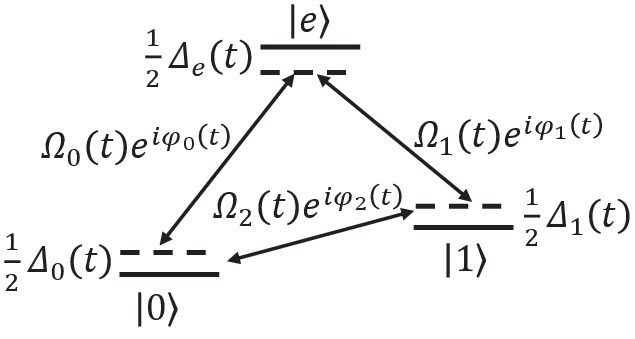}
\caption{Sketch of a general three-level system under control, in which the transitions $|0\rangle\leftrightarrow|e\rangle$, $|1\rangle\leftrightarrow|e\rangle$ and $|0\rangle\leftrightarrow|1\rangle$ are driven by the off-resonant driving fields.}\label{model}
\end{figure}

This section contributes to the construction of universal paths, either adiabatic or nonadiabatic, for a general three-level system driven by three off-resonant driving fields, as shown in Fig~\ref{model}. It serves as a pedagogical model to illustrate the dynamical-correction condition in Eq.~(\ref{OptGeneral}), as presented in the previous section. The transitions among the three levels are induced by off-resonant laser fields. In particular, the transition $|n\rangle\leftrightarrow|e\rangle$, with $n=0,1$, is driven by the field with the time-dependent Rabi frequency $\Omega_n(t)$, the time-dependent phase $\varphi_n(t)$, and detuning $[\Delta_e(t)-\Delta_n(t)]/2$. The transition $|0\rangle\leftrightarrow|1\rangle$ is driven by the field $\Omega_2(t)$ with the phase $\varphi_2(t)$ and detuning $[\Delta_1(t)-\Delta_0(t)]/2$. In experiments, the transition $|0\rangle\leftrightarrow|1\rangle$ of the superconducting transmon qubit can be achieved with a two-photon process~\cite{Antti2019Superadiabatic}. Then the ideal Hamiltonian reads
\begin{equation}\label{Ham1}
\begin{aligned}
H_0(t)&=\frac{1}{2}\left[\Delta_e(t)|e\rangle\langle e|+\Delta_1(t)|1\rangle\langle 1|+\Delta_0(t)|0\rangle\langle 0|\right]\\ &+\Big[\Omega_0(t)e^{i\varphi_0(t)}|e\rangle\langle 0|+\Omega_1(t)e^{i\varphi_1(t)}|e\rangle\langle 1|\\
&+\Omega_2(t)e^{i\varphi_2(t)}|1\rangle\langle0|+{\rm H.c.}\Big].
\end{aligned}
\end{equation}

To exactly solve the Schr\"odinger equation under the time-dependent Hamiltonian in Eq.~(\ref{Ham1}), one can resort to the algorithm described within the ancillary picture in Sec.~\ref{Ideal}. The basis states spanning the ancillary picture~\cite{Jin2024Shortcut,Jin2024Entangling} can be chosen to be 
\begin{subequations}\label{Anci}
\begin{align}
|\mu_1(t)\rangle&=\cos\theta(t)e^{i\frac{\alpha_0(t)}{2}}|0\rangle-\sin\theta(t)e^{-i\frac{\alpha_0(t)}{2}}|1\rangle, \label{mu1} \\
|\mu_2(t)\rangle&=\cos\phi(t)e^{i\frac{\alpha(t)}{2}}|b(t)\rangle-\sin\phi(t)e^{-i\frac{\alpha(t)}{2}}|e\rangle, \label{mu2} \\
|\mu_3(t)\rangle&=\sin\phi(t)e^{i\frac{\alpha(t)}{2}}|b(t)\rangle+\cos\phi(t)e^{-i\frac{\alpha(t)}{2}}|e\rangle, \label{mu3}
\end{align}
\end{subequations}
where $b(t)\equiv\sin\theta(t)e^{i\alpha_0(t)/2}|0\rangle+\cos\theta(t)e^{-i\alpha_0(t)/2}|1\rangle$ is orthonormal to $|\mu_1(t)\rangle$. The undetermined parameters $\theta(t)$ and $\phi(t)$ are relevant to the population transfer, and $\alpha_0(t)$ and $\alpha(t)$ manipulate the local phases.

These ancillary basis states in Eq.~(\ref{Anci}) apply to any three-level systems and could be easily extended to larger systems~\cite{Jin2024Entangling}. Substituting them into the von Neumann equation~(\ref{von}) with the Hamiltonian~(\ref{Ham1}), the detunings and the Rabi-frequencies in the laboratory frame are found to be
\begin{equation}\label{detuning}
\begin{aligned}
\Delta_e(t)&=\Delta(t),\\
\Delta_1(t)&=-\Delta(t)\cos^2\theta(t)+\Delta_a(t),\\
\Delta_0(t)&=-\Delta(t)\sin^2\theta(t)-\Delta_a(t),
\end{aligned}
\end{equation}
and
\begin{equation}\label{Rabifrequency}
\begin{aligned}
\Omega_0(t)e^{i\varphi_0(t)}&=\Omega(t)\sin\theta(t)e^{-i\frac{\alpha_0(t)}{2}},\\
\Omega_1(t)e^{i\varphi_1(t)}&=\Omega(t)\cos\theta(t)e^{i\frac{\alpha_0(t)}{2}},\\
\Omega_2(t)e^{i\varphi_2(t)}&=\Omega_a(t)-\frac{1}{2}\Delta(t)\sin\theta(t)\cos\theta(t)e^{-i\alpha_0(t)},
\end{aligned}
\end{equation}
respectively, where the scaling detunings $\Delta(t)$ and $\Delta_a(t)$ and the scaling Rabi frequencies $\Omega(t)$ and $\Omega_a(t)$ satisfy the following conditions
\begin{equation}\label{ConditionMu0}
\begin{aligned}
&\Delta_a(t)=\dot{\alpha}_0(t)+2\Omega_a(t)\cot2\theta(t)\cos\alpha_0(t),\\
&\Omega_a(t)=-\frac{\dot{\theta}(t)}{\sin\alpha_0(t)},\\
&\Delta(t)=\dot{\alpha}(t)+2\Omega(t)\cot2\phi(t)\cos\alpha(t)+\Omega_a(t)\frac{\cos\alpha_0(t)}{\sin2\theta(t)},\\
&\Omega(t)=-\frac{\dot{\phi}(t)}{\sin\alpha(t)}.
\end{aligned}
\end{equation}
Under Eqs.~(\ref{detuning}), (\ref{Rabifrequency}), and (\ref{ConditionMu0}), all the three $|\mu_k(t)\rangle$'s can be activated as universal nonadiabatic paths. Also, they construct the exact solution to the time-dependent Schr\"odinger equation, i.e., the full-rank evolution operator in Eq.~(\ref{U0}) with $K=3$. In the absence of systematic errors, the time-evolution operator for the three-level system~(\ref{Ham1}) can be written as
\begin{equation}\label{Uthree}
U(t,0)=\sum_{k=1}^3e^{if_k(t)}|\mu_k(t)\rangle\langle\mu_k(0)|,
\end{equation}
where the global phases can be expressed as
\begin{equation}\label{globalphase}
\begin{aligned}
&\dot{f}_1(t)=\Omega_a(t)\frac{\cos\alpha_0(t)}{\sin2\theta(t)}=-\dot{\theta}(t)\frac{\cot\alpha_0(t)}{\sin2\theta(t)},\\
&\dot{f}_2(t)=\dot{f}(t)-\frac{1}{2}\dot{f}_1(t), \\
&\dot{f}_3(t)=-\dot{f}(t)-\frac{1}{2}\dot{f}_1(t)
\end{aligned}
\end{equation}
with
\begin{equation}\label{globalphasef}
\dot{f}(t)=\Omega(t)\frac{\cos\alpha(t)}{\sin2\phi(t)}=-\dot{\phi}(t)\frac{\cot\alpha(t)}{\sin2\phi(t)}.
\end{equation}

A well-known special case occurs when the global phase of $|\mu_1(t)\rangle$ is assumed to be constant, i.e., $\dot{f}_1(t)=0$ indicating $\alpha_0(t)=\pi/2$ due to Eq.~(\ref{globalphase}), which is exactly the parallel-transport condition for the single path $|\mu_1(t)\rangle$ in the absence of errors~\cite{Ming2022Experimental,Andr2022Dark,Jin2024Geometric,Sjoqvist2012Nonadiabatic,Liu2019Plug,Jing2017Non,
Zheng2016Comparison}. Under this assumption, the global phase for the other two paths can be written as
\begin{equation}\label{globalphase1}
\dot{f}_2(t)=\dot{f}(t), \quad \dot{f}_3(t)=-\dot{f}(t),
\end{equation}
due to Eq.~(\ref{globalphase}). In Sec.~\ref{NonidealThree}, we will show that $\dot{f}(t)\ne0$ can be used to fulfill our dynamical-correction condition in Eq.~(\ref{OptGeneral}) and the adverse effects arising from the systematic errors can be significantly suppressed.

With no loss of generality, one can choose $\theta(t)$, $\phi(t)$, $f_1(t)$, and $f(t)$ to be independent variables in realtime control. Then with Eqs.~(\ref{globalphase}) and (\ref{globalphasef}), Eq.~(\ref{ConditionMu0}) can be rewritten as
\begin{equation}\label{Mu0Inverse}
\begin{aligned}
\Delta_a(t)&=\dot{\alpha}_0(t)+2\dot{f}_1(t)\cos2\theta(t),\\
\dot{\alpha}_0(t)&=-\frac{\ddot{\theta}\dot{f}_1\sin2\theta-\ddot{f}_1\dot{\theta}\sin2\theta
-2\dot{f}_1\dot{\theta}^2\cos2\theta}{\dot{f}_1^2\sin^22\theta+\dot{\theta}^2},\\
\Omega_a(t)&=-\sqrt{\dot{\theta}(t)^2+\dot{f}_1(t)^2\sin^22\theta(t)},
\end{aligned}
\end{equation}
and
\begin{equation}\label{Mu12Inverse}
\begin{aligned}
\Delta(t)&=\dot{\alpha}(t)+2\dot{f}(t)\cos2\phi(t)+\dot{f}_1(t),\\
\dot{\alpha}(t)&=-\frac{\ddot{\phi}\dot{f}\sin2\phi-\ddot{f}\dot{\phi}\sin2\phi
-2\dot{f}\dot{\phi}^2\cos2\phi}{\dot{f}^2\sin^22\phi+\dot{\phi}^2},\\
\Omega(t)&=-\sqrt{\dot{\phi}(t)^2+\dot{f}(t)^2\sin^22\phi(t)}.
\end{aligned}
\end{equation}
We find that the system evolution along the path $|\mu_1(t)\rangle$ is determined by the parameters $\theta(t)$ and $f_1(t)$, which can be controlled by $\Delta_a(t)$ and $\Omega_a(t)$ in Eq.~(\ref{Mu0Inverse}). Similarly, the setting of $\phi(t)$, $f_1(t)$, and $f(t)$ determines the evolution along the path $|\mu_2(t)\rangle$ or $|\mu_3(t)\rangle$, which is controlled by $\Delta(t)$ and $\Omega(t)$ in Eq.~(\ref{Mu12Inverse}). Equation~(\ref{Mu12Inverse}) has a form similar to Eq.~(\ref{Mu0Inverse}) due to the fact that the transition between the states $|b(t)\rangle$ and $|e\rangle$ is equivalent to that in an effective two-level system.

We take the unidirectional population transfer $|0\rangle\rightarrow|e\rangle\rightarrow|1\rangle$ along the path $|\mu_2(t)\rangle$ as an example. For this task, $\phi(t)$ and $\theta(t)$ are required to satisfy the boundary conditions $\phi(0)=0$, $\phi(T/2)=\pi/2$, $\phi(T)=\pi$, $\theta(0)=\pi/2$, and $\theta(T)=\pi$, where $T$ is a running period. Accordingly, one can set
\begin{equation}\label{ThePhiMu0}
\phi(t)=\frac{\pi t}{T}, \quad \theta(t)=\frac{\pi}{2}+\frac{\phi(t)}{2},
\end{equation}
with which the initial population on level $|0\rangle$ can be completely transferred to level $|e\rangle$ when $t=T/2$ and then to level $|1\rangle$ when $t=T$.

\subsection{Nonideal situation}\label{NonidealThree}

Systematic errors due to inherent parameter variations during the time evolution give rise to unwanted transitions among different paths. This is a central challenge in quantum control. In Eq.~(\ref{SchNon}), the non-ideal system dynamics for the model in Fig.~\ref{model} is controlled by the full Hamiltonian $H(t)=H_0(t)+\epsilon H_1(t)$. Here $H_0(t)$ is the ideal Hamiltonian in Eq.~(\ref{Ham1}) and $H_1(t)$ is the error Hamiltonian formulated as part of $H_0(t)$. $H_1(t)$ can be roughly classified into two categories: (1) the global errors representing correlated parameter fluctuations, and (2) the local errors describing individual parameter variations. From a pedagogical perspective, the global error Hamiltonian can be chosen to be
\begin{equation}\label{globalerror}
\begin{aligned}
H_1(t)&=\Omega_0(t)e^{i\varphi_0(t)}|e\rangle\langle 0|+\Omega_1(t)e^{i\varphi_1(t)}|e\rangle\langle 1|\\
&+\Omega_2(t)e^{i\varphi_2(t)}|1\rangle\langle0|+{\rm H.c.},
\end{aligned}
\end{equation}
which indicates the fully correlated or synchronous deviation or dispersion in all three driving-fields' Rabi frequencies and phases. On the other hand, the local systematic error can be chosen as
\begin{equation}\label{delta1}
H_1(t)=\frac{1}{2}\Delta_1(t)|1\rangle\langle 1|+\left[\Omega_0(t)e^{i\varphi_0(t)}|e\rangle\langle0|+{\rm H.c.}\right],
\end{equation}
which involves only the fluctuations in the driving field about the transition $|e\rangle\leftrightarrow|0\rangle$ and the detuning variation of level $|1\rangle$. Our dynamical-correction mechanism achieves comprehensive suppression of systematic imperfections by demonstrating robustness against both global error in Eq.~(\ref{globalerror}) and local error in Eq.~(\ref{delta1}).

For the three-level system in the presence of the global error in Eq.~(\ref{globalerror}), the matrix elements of the error rotation operator $\mathcal{M}_{kn}(t)$ in Eq.~(\ref{Convension}) are found to be
\begin{equation}\label{errorTh}
\begin{aligned}
&|\mathcal{M}_{12}(t)|\\
&=\left|\int_0^t\frac{1}{4}\left[\dot{\alpha}+2\dot{f}\cos(2\phi)\right]\sin(4\theta)\cos\phi e^{i(\frac{\alpha}{2}+f)}dt_1\right|,\\
&|\mathcal{M}_{13}(t)|\\
&=\left|\int_0^t\frac{1}{4}\left[\dot{\alpha}+2\dot{f}\cos(2\phi)\right]\sin(4\theta)\sin\phi e^{i(\frac{\alpha}{2}+f)}dt_1\right|,\\
&|\mathcal{M}_{23}(t)|=\Big|\int_0^t\Big\{\dot{f}+\frac{1}{4}\left[\dot{\alpha}-2\dot{f}\cos(2\phi)\right]\sin^2(2\theta)\\
&+i\dot{\phi}\Big\}\sin(2\phi)e^{-i2f}dt_1\Big|
\end{aligned}
\end{equation}
using Eqs.~(\ref{globalphase1}), (\ref{Mu0Inverse}), (\ref{Mu12Inverse}) and (\ref{globalerror}). Here $\dot{f}(t)$ and $\dot{\alpha}(t)$ are determined by Eqs.~(\ref{globalphasef}) and (\ref{Mu12Inverse}), respectively. It is found that $|\dot{\alpha}(t)|$ will approach zero when $|\dot{f}(t)|$ is sufficiently small.

Focusing on the kernel in the time integral, one can see that Eq.~(\ref{errorTh}) provides a unified perspective of the essences and constraints of various quantum control protocols, such as the stimulated Raman adiabatic passage~\cite{Vitanov2017Stiumlated} and the nonadiabatic holonomic transformation~\cite{Sjoqvist2012Nonadiabatic,Liu2019Plug}. In the stimulated Raman adiabatic passage, the main Hamiltonian~(\ref{Ham1}) itself is slowly varying with time, i.e., $\dot{\Delta}_a(t),\dot{\Omega}_a(t),\dot{\Delta}(t),\dot{\Omega}(t)\approx0$. According to Eqs.~(\ref{Mu0Inverse}) and (\ref{Mu12Inverse}) and their time derivatives, this suggests that $\dot{\theta}_0(t),\dot{\alpha}_0(t),\dot{f}_1(t),\dot{\phi}(t),\dot{\alpha}(t),\dot{f}(t)\approx0$. The off-diagonal elements $|\mathcal{M}_{kn}|$, with $k\neq n\in\{1,2,3\}$, can then become vanishing under such an adiabatic condition. In essence, the adiabatic control is the limiting case to the correction mechanism (1) such that the time integral is divided into infinite segments and the error can be self-corrected in each segment. The system, however, is sensitive to decoherence under the much extended exposure to the environment and the adiabatic path is expected to fail. Under the parallel-transport condition for all three paths in holonomic transformation~\cite{Sjoqvist2012Nonadiabatic,Liu2019Plug}, i.e., $\dot{f}(t)=0$, which enforces $\dot{\alpha}(t)=0$ via Eq.~(\ref{Mu12Inverse}), $|\mathcal{M}_{12}|$ and $|\mathcal{M}_{13}|$ in Eq.~(\ref{errorTh}) will be vanishing. $|\mathcal{M}_{23}|$, however, cannot be canceled due to the term proportional to $\dot{\phi}(t)$. Thus it justifies that the holonomic transformation is significantly sensitive to the systematic errors~\cite{Jing2017Non,Zheng2016Comparison}. Shortcuts to adiabaticity omitted the role of the path-dependent global phases~\cite{Vitanov2017Stiumlated} in state transfer. Nevertheless, we find that they can be used as an explicit degree of freedom for robust quantum control.

In sharp contrast to the strict condition $\dot{f}(t)=0$ for the full-rank holonomic transformation, our dynamical-correction condition in Eq.~(\ref{OptGeneral}) can be attributed to a sufficiently large $|\dot{f}(t)|$ under the assumption that $\dot{f}_1(t)=0$. With no loss of generality, we can simply set
\begin{equation}\label{Simplefdot}
f(t)=\lambda\phi(t),
\end{equation}
or
\begin{equation}\label{Simplef}
\dot{f}(t)=\lambda\dot{\phi}(t),
\end{equation}
where $\phi(t)$ can be an arbitrary smooth function of time. Keeping in mind that $\phi(t)$, $f(t)$, $\theta(t)$, and $f_1(t)$ are chosen to be independent variables in our theory, we can determine a unique evolution for the three-level system by putting Eqs.~(\ref{globalphasef}) and (\ref{Simplefdot}) or (\ref{Simplef}) together. With Eqs.~(\ref{detuning}), (\ref{Rabifrequency}), (\ref{Mu0Inverse}), (\ref{Mu12Inverse}), and (\ref{Simplef}), the detunings, Rabi frequencies, and phases in the original Hamiltonian~(\ref{Ham1}) can be found to be
\begin{equation}\label{detuori}
\begin{aligned}
\Delta_e(t)&=\dot{\alpha}+2\lambda\dot{\phi}\cos(2\phi),\\
\Delta_1(t)&=-\left[\dot{\alpha}+2\lambda\dot{\phi}\cos(2\phi)\right]\cos^2\theta,\\
\Delta_0(t)&=-\left[\dot{\alpha}+2\lambda\dot{\phi}\cos(2\phi)\right]\sin^2\theta,
\end{aligned}
\end{equation}
and
\begin{equation}\label{Rabiri}
\begin{aligned}
\Omega_0(t)e^{i\varphi_0(t)}&=-|\dot{\phi}|\sqrt{1+\lambda^2\sin^2(2\phi)}\sin\theta e^{-i\alpha_0/2},\\
\Omega_1(t)e^{i\varphi_1(t)}&=-|\dot{\phi}|\sqrt{1+\lambda^2\sin^2(2\phi)}\cos\theta e^{i\alpha_0/2},\\
\Omega_2(t)e^{i\varphi_2(t)}&=-\dot{\theta}-\frac{1}{4}\left[\dot{\alpha}+2\lambda\dot{\phi}\cos(2\phi)\right]
\sin(2\theta)e^{-i\alpha_0},
\end{aligned}
\end{equation}
where $\theta(t)$ and $\phi(t)$ can be arbitrarily chosen as long as their variation rates are much smaller than $|\dot{f}(t)|$. Here we use the definitions for $\theta(t)$ and $\phi(t)$ in Eq.~(\ref{ThePhiMu0}). The coefficient $\lambda$ scales the relative magnitude between $\dot{f}(t)$ and $\dot{\phi}(t)$. When $\lambda=0$, Eqs.~(\ref{detuori}) and (\ref{Rabiri}) are found to be consistent with the state-transfer protocol using a full-rank holonomic transformation in Ref.~\cite{Jin2024Shortcut}.

The power of the correction condition~(\ref{Simplef}) can be analytically illustrated through the technique of integration by parts. In the setting of $\theta(t)$, $\phi(t)$, and $f(t)$ in Eqs.~(\ref{ThePhiMu0}) and (\ref{Simplef}), $\dot{\alpha}(t)$ can be estimated by Eq.~(\ref{Mu12Inverse}) as
\begin{equation}\label{dalpha}
\begin{aligned}
&\left|\frac{\dot{\alpha}(t)}{2}\right|=\left|\frac{\dot{f}\dot{\phi}^2\cos(2\phi)}
{\dot{f}^2\sin^2(2\phi)+\dot{\phi}^2}\right|=\left|\frac{\dot{f}\cos(2\phi)}{(\dot{f}^2/\dot{\phi}^2)\sin^2(2\phi)+1}\right|\\
&=\left|\frac{\dot{f}\cos(2\phi)}{\lambda^2\sin^2(2\phi)+1}\right|\leq\left|\dot{f}(t)\right|,
\end{aligned}
\end{equation}
where the last equality holds in the case of $\phi(t)=k\pi$, $k\in Z$. Using Eq.~(\ref{dalpha}) and repeated integrations by parts, $|\mathcal{M}_{12}(t)|$ in Eq.~(\ref{errorTh}) are bounded by
\begin{equation}\label{errorThReduce}
\begin{aligned}
&|\mathcal{M}_{12}(T)|=\left|\int_0^T\frac{\dot{\alpha}+2\dot{f}\cos(2\phi)}{4\dot{f}}\sin(4\theta)\cos\phi e^{i\frac{\alpha}{2}}de^{if}\right|\\
&\leq\left|\int_0^TF_1(t)de^{if}\right|=\left|\int_0^Te^{if}\left[F_{\theta,\phi}+i\frac{\dot{\alpha}}{2}F_1(t)\right]dt\right|\\
&\leq\left|\int_0^T\frac{F_{\theta,\phi}}{\dot{f}}de^{if}\right|+\left|\int_0^T\frac{\dot{\alpha}}{2\dot{f}}F_1de^{if}\right|=\left|\int_0^T\frac{F_{\theta,\phi}}{\dot{f}}de^{if}\right|\\
&+\left|\int_0^T\frac{\dot{\alpha}}{2\dot{f}}\left(\frac{F_{\theta,\phi}}{i\dot{f}}+\frac{\dot{\alpha}}{2\dot{f}}F_1\right)de^{if}\right|\\
&=\cdots=\left|\int_0^T\frac{F_{\theta,\phi}}{\dot{f}}de^{if}\right|+\big|\sum_{k=1}^\infty\int_0^T\Big(\frac{\dot{\alpha}}{2\dot{f}}\Big)^k\frac{F_{\theta,\phi}}{i\dot{f}}de^{if}\\
&+\lim_{k\rightarrow\infty}\int_0^T\Big(\frac{\dot{\alpha}}{2\dot{f}}\Big)^kF_1(t)de^{if}\big|\\
&\le\sum_{k=0}^{\infty}\left|\int_0^Te^{if}d
\left[\left(\frac{\dot{\alpha}}{2\dot{f}}\right)^k\frac{F_{\theta,\phi}}{\dot{f}}\right]\right|\\
&+\lim_{k\rightarrow\infty}\left|\int_0^Te^{if}\left(\frac{\dot{\alpha}}{2\dot{f}}\right)^kF_1(t)dt\right|,
\end{aligned}
\end{equation}
where
\begin{equation}\label{FG}
\begin{aligned}
&F_1(t)\equiv\sin(4\theta)\cos(2\phi)\cos\phi e^{i\frac{\alpha}{2}}, \\
&F_{\theta,\phi}\equiv\frac{d}{dt}[\sin(4\theta)\cos(2\phi)\cos\phi]e^{i\frac{\alpha}{2}}.
\end{aligned}
\end{equation}
The third equivalence in Eq.~(\ref{errorThReduce}) uses the boundary condition $F_1(T)=0$ and the relation $dF_1(t)/dt=F_{\theta,\phi}+i(\dot{\alpha}/2)F_1(t)$ according to Eq.~(\ref{FG}).

Both components in the final result of Eq.~(\ref{errorThReduce}) consist of a pseudo-periodical function $e^{if(t)}$. They will approach zero when the global phase $f(t)$ is rapidly varying with time. Consequently, $|\mathcal{M}_{12}(T)|$ vanishes in a finite interval of $T$. Similar reasoning applies to the vanishing of the second element $|\mathcal{M}_{13}(T)|$ in Eq.~(\ref{errorTh}).

As for the third element $|\mathcal{M}_{23}(T)|$, we have
\begin{equation}\label{errorThLe}
|\mathcal{M}_{23}(T)|\le\left|\int_0^TF_2(t)de^{if}\right|
=\left|\int_0^Te^{if}d\left(\frac{\dot{F}_2(t)}{\dot{f}}\right)\right|,
\end{equation}
where
\begin{equation}
  F_2(t)=\sin(2\phi)-\sin(4\phi)\sin^2(2\theta)/4+i\sin(2\phi)/\lambda.
\end{equation}
It is straightforward to verify the vanishing of $|\mathcal{M}_{23}(T)|$ in the presence of a sufficiently large $|\lambda|$ in Eq.~(\ref{Simplef}).

The performance of our protocol can be evaluated by the fidelity $\mathcal{F}(T)\equiv|\langle\psi(T)|1\rangle|^2$, where the system wave function $|\psi(t=T)\rangle$ is obtained with the time-dependent Schr\"odinger equation $i\partial|\psi(t)\rangle/\partial t=[H_0(t)+\epsilon H_1(t)]|\psi(t)\rangle$ with the ideal Hamiltonian $H_0(t)$~(\ref{Ham1}) and the error Hamiltonian $H_1(t)$ in either Eq.~(\ref{globalerror}) or (\ref{delta1}). However, in the error-free dynamics, the system initially prepared as $|\psi(0)\rangle=|0\rangle$ would evolve to the targets $|\psi(T/2)\rangle=|e\rangle$ and $|\psi(T)\rangle=|1\rangle$ along the nonadiabatic path $|\mu_2(t)\rangle$ for the unidirectional population transfer.

\begin{figure}[htbp]
\centering
\includegraphics[width=0.85\linewidth]{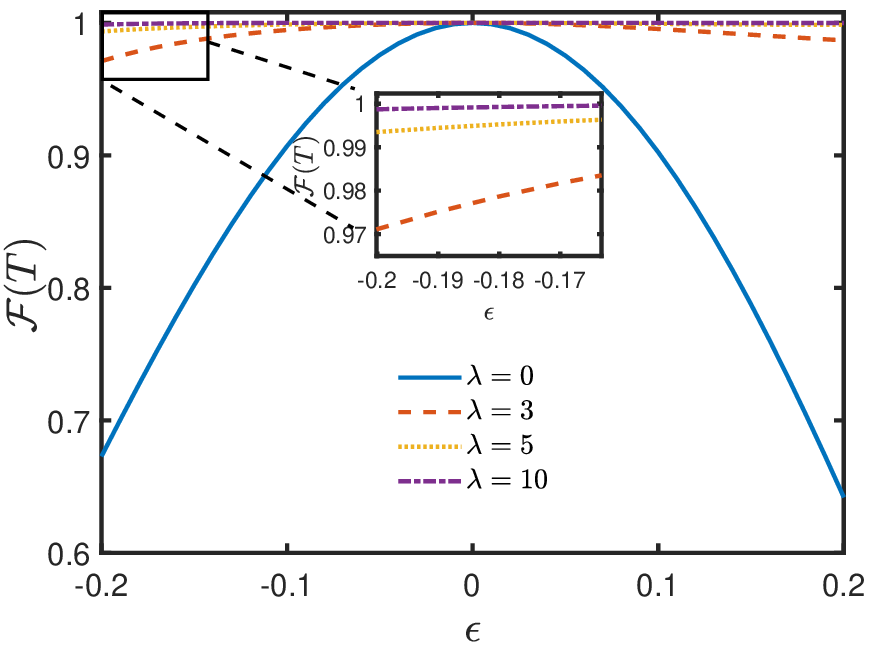}
\caption{Fidelity $\mathcal{F}(T)$ versus the global error $\epsilon$ for the complete population transfer $|0\rangle\rightarrow|e\rangle\rightarrow|1\rangle$ along the path $|\mu_2(t)\rangle$ in Eq.~(\ref{mu2}). The detuning, the Rabi frequencies, and the phases are set from Eqs.~(\ref{detuning}), (\ref{Rabifrequency}), (\ref{Mu0Inverse}), and (\ref{Mu12Inverse}) with $\theta(t)$ and $\phi(t)$ in Eq.~(\ref{ThePhiMu0}), $f(t)$ in Eq.~(\ref{Simplef}) with various coefficients $\lambda$, and $f_1(t)=\pi/2$.}\label{Pe_dfdphi}
\end{figure}

In Fig.~\ref{Pe_dfdphi}, for various coefficients $\lambda$, the fidelity at $t=T$ is demonstrated versus the error magnitude $\epsilon$ for the global error Hamiltonian~(\ref{globalerror}). We choose $|\mu_2(t)\rangle$ in Eq.~(\ref{mu2}) to be evolution path that could be controlled by the global phase $f_2(t)$ with $|f_2(t)|=|f(t)|$. Figure.~\ref{Pe_dfdphi} shows that under the conventional parallel-transport condition, i.e., $\lambda=0$ or $\dot{f}(t)=0$, the fidelity is sensitive to the systematic error. In particular, $\mathcal{F}=0.673$ when $\epsilon=-0.2$, $\mathcal{F}=0.910$ when $\epsilon=-0.1$, $\mathcal{F}=0.902$ when $\epsilon=0.1$, and $\mathcal{F}=0.642$ when $\epsilon=0.2$. In comparison to a constant global phase $f$, the rapidly-varying global phase $f(t)$ can efficiently correct the adverse effects arising from the global error. With $\lambda=\dot{f}(t)/\dot{\phi}(t)=3$, the fidelity $\mathcal{F}$ is above $0.971$ in the range of $\epsilon\in[-0.2, 0.2]$. With $\lambda=5$, the lower-bound of the fidelity is enhanced to about $0.993$. And when $\lambda=10$, the fidelity is maintained as unit even if the error magnitude is as large as $20\%$ of the ideal Hamiltonian.

Next we examine the validity of our dynamical error-correction mechanism in Eq.~(\ref{OptGeneral}) in the presence of the local error Hamiltonian~(\ref{delta1}). It describes a type of error in which both detuning $\Delta_1(t)$ and the Rabi frequency $\Omega_0(t)$ deviate from the original setting. In a way similar to the global error, using the conditions in Eqs.~(\ref{globalphase1}), (\ref{Mu0Inverse}), (\ref{Mu12Inverse}), (\ref{delta1}) and (\ref{Simplef}), and the boundary conditions of $\theta(t)$ and $\phi(t)$, $|\mathcal{M}_{12}(t)|$ defined in Eq.~(\ref{Convension}) is found to be upper-bounded by
\begin{equation}\label{errorNoncom}
\begin{aligned}
&|\mathcal{M}_{12}(T)|\le\left|\int_0^TG_1(t)de^{if}\right|+\left|\int_0^TG_2(t)de^{if}\right|\\
&\le\sum_{m=1}^2\sum_{k=0}^{\infty}\left|\int_0^Te^{if}d\left[\left(\frac{\dot{\alpha}}{2\dot{f}}\right)^k
\frac{G_{\theta,\phi}^{(m)}}{\dot{f}}\right]\right|\\
&+\sum_{m=1}^2\lim_{k\rightarrow\infty}\left|\int_0^Te^{if}\left(\frac{\dot{\alpha}}{2\dot{f}}\right)^kG_m(t)dt\right|,
\end{aligned}
\end{equation}
where
\begin{equation}\label{FGNon}
\begin{aligned}
&G_1(t)\equiv\sin(2\theta)\cos^2\theta\cos(2\phi)\cos\phi e^{i\frac{\alpha}{2}}, \\
&G_2(t)\equiv\sin\theta\cos\theta\sin\phi e^{-i\frac{\alpha}{2}},\\
&G_{\theta,\phi}^{(m)}=\dot{G}_{m}(t)-i\frac{\dot{\alpha}}{2}G_m(t).
\end{aligned}
\end{equation}
Similar to Eqs.~(\ref{errorThReduce}) and (\ref{errorThLe}), the integrals in Eq.~(\ref{errorNoncom}) are found to be vanishing with a large $|\dot{f}(t)|$. Similarly, the other two off-diagonal elements $|\mathcal{M}_{13}(t)|$ and $|\mathcal{M}_{23}(t)|$ defined in Eq.~(\ref{Convension}) are found to approach zero when $|\lambda|$ is sufficiently large.

\begin{figure}[htbp]
\centering
\includegraphics[width=0.85\linewidth]{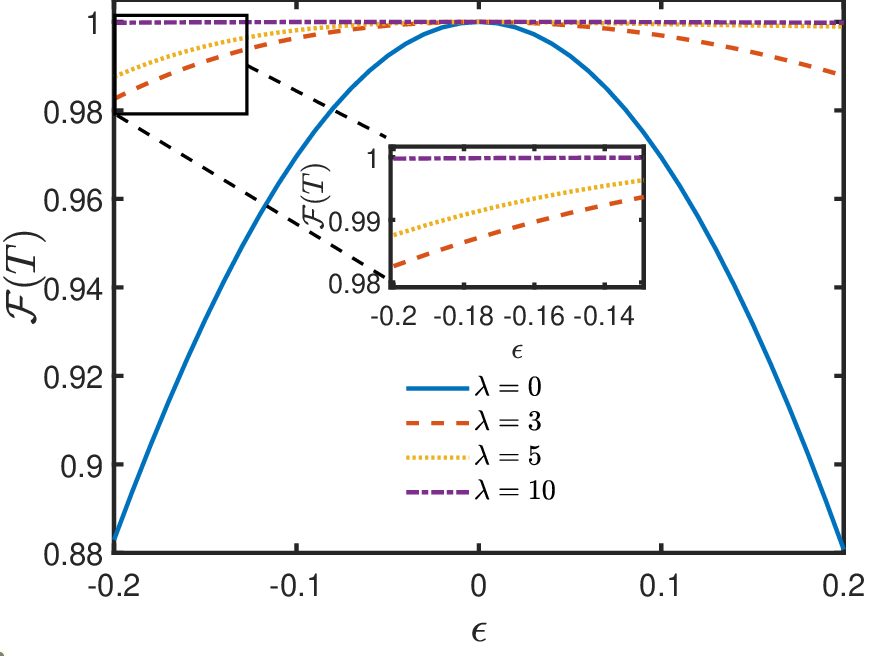}
\caption{Fidelity $\mathcal{F}(T)$ versus the local error $\epsilon$ for the complete population transfer $|0\rangle\rightarrow|e\rangle\rightarrow|1\rangle$ along the path $|\mu_2(t)\rangle$ in Eq.~(\ref{mu2}). The parameters are set same as Fig.~\ref{Pe_dfdphi}.}\label{P1_dfdphi}
\end{figure}

Under various coefficients $\lambda$'s, Fig.~\ref{P1_dfdphi} demonstrates the fidelity at $t=T$ versus the error magnitude $\epsilon$. The chosen path $|\mu_2(t)\rangle$ embedded in our dynamical-correction mechanism is found to be robust against the local error as well, in comparison to that under the parallel-transport condition. Particularly, when the evolution satisfies the parallel-transport condition $\lambda=0$, the fidelity is about $\mathcal{F}=0.883$ when $\epsilon=\pm0.2$. In contrast, the fidelity is lower bounded by $\mathcal{F}=0.983$ when $\lambda=3$ and by $0.988$ when $\lambda=5$ in the range of $\epsilon\in[-0.2, 0.2]$. When $\lambda=10$, the fidelity remains almost unit irrespective of the error magnitude.

\section{Cyclic population transfer with error correction}\label{PopErr}

In this section, we intend to show the power of our dynamical error-correction protocol in the cyclic population transfer of the general three-level system in Fig.~\ref{model}. With a fixed and exaggerated error magnitude $\epsilon=-0.2$, the calculation of the state population further confirms the validity of the universal correction condition in Eq.~(\ref{OptGeneral}) or its simplified and specific version in Eq.~(\ref{Simplef}).

\begin{figure}[htbp]
\centering
\includegraphics[width=0.85\linewidth]{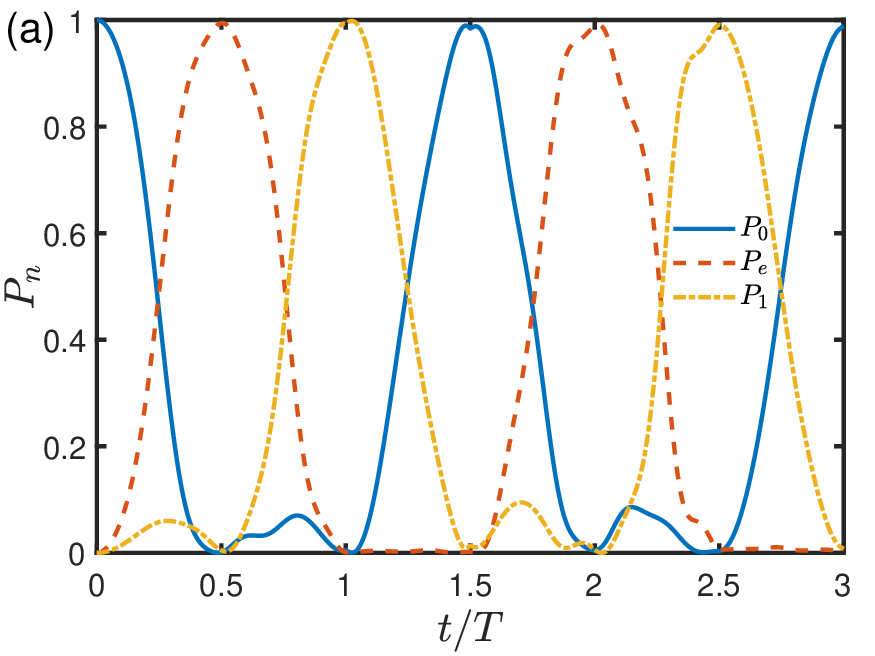}
\includegraphics[width=0.85\linewidth]{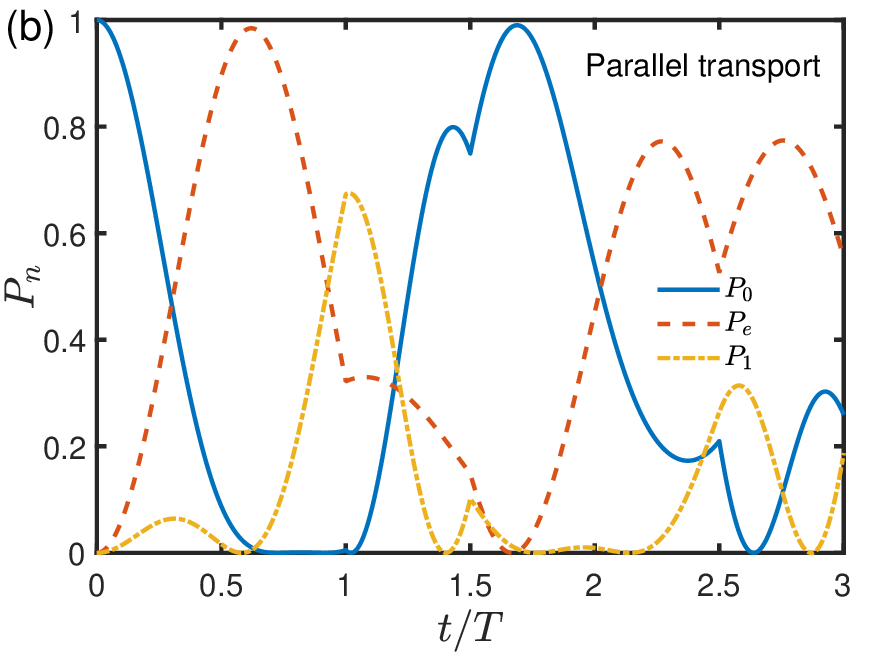}
\caption{State populations $P_n(t)$, $n=0,1,e$, under the global error in Eq.~(\ref{globalerror}) with $\epsilon=-0.2$, versus the evolution time $t/T$ during the cyclic population transfer. The scaling coefficient $\lambda$ in Eq.~(\ref{Simplef}) for the global phase $f(t)$ is fixed as (a) $\lambda=5$ and (b) $\lambda=0$. The detuning, the Rabi frequencies, and the phases are set from Eqs.~(\ref{detuning}), (\ref{Rabifrequency}), (\ref{Mu0Inverse}), and (\ref{Mu12Inverse}) with $\theta(t)$ and $\phi(t)$ in Eqs.~(\ref{boundu2second}) and (\ref{boundu0second}), and $f_1(t)=\pi/2$.}\label{CycRabi}
\end{figure}

A complete loop of the cyclic population transfer $|0\rangle\rightarrow|e\rangle\rightarrow|1\rangle\rightarrow|0\rangle$ can be divided into two stages, i.e., (1) $|0\rangle\rightarrow|e\rangle\rightarrow|1\rangle$ lasting $T$ and (2) $|1\rangle\rightarrow|0\rangle$ lasting $T/2$. In stage 1, one can employ the universal path $|\mu_2(t)\rangle$, in which $\theta(t)$ and $\phi(t)$ are set according to Eq.~(\ref{ThePhiMu0}). The system along the path starts from $|0\rangle$ and becomes $|e\rangle$ when $t=T/2$ and $|1\rangle$ when $t=T$. Stage 2 employs the path $|\mu_1(t)\rangle$ in Eq.~(\ref{mu1}), where the boundary conditions of $\theta(t)$ are $\theta(T)=\pi/2$ and $\theta(3T/2)=\pi$. Thus on stage 2 one can simply set
\begin{equation}\label{CycSg2Thephi}
\phi(t)=\frac{\pi t}{T}, \quad \theta(t)=\phi(t)-\frac{\pi}{2}.
\end{equation}
Then the population in $|1\rangle$ can be completely transferred to $|0\rangle$ when $t=3T/2$.

In general, the $k$th loop of the cyclic population transfer during $t\in[3(k-1)T/2, 3kT/2]$, $k\geq1$, can be divided into two stages. In the first stage via the path $|\mu_2(t)\rangle$, the parameters can be set as
\begin{equation}\label{boundu2second}
\phi(t)=\frac{\pi\left[2t-3(k-1)T\right]}{2T}, \quad \theta(t)=\frac{\pi}{2}+\frac{\phi(t)}{2},
\end{equation}
In the second one via $|\mu_1(t)\rangle$, we have
\begin{equation}\label{boundu0second}
\phi(t)=\frac{\pi\left[2t-3(k-1)T\right]}{2T}, \quad \theta(t)=\phi(t)-\frac{\pi}{2}.
\end{equation}
When $t=3kT/2$, the population of the system should be completely transferred back to the initial state $|0\rangle$. The presence of systematic errors, i.e., $\epsilon\ne0$, will induce undesirable transitions among various paths during the cyclic population transfer. To demonstrate the correction effect of the global phase $f(t)$ set to Eq.~(\ref{Simplef}), the performance of our protocol for the cyclic population transfer can be evaluated by the level populations $P_n(t)\equiv\langle n|\psi(t)\rangle\langle\psi(t)|n\rangle$, with $n=0,1,e$. Note the fidelity $\mathcal{F}(T)$ defined in Sec.~\ref{Optimize} is actually a special case of $P_n$, i.e., $\mathcal{F}(T)=P_1(t=T)$.

\begin{figure}[htbp]
\centering
\includegraphics[width=0.85\linewidth]{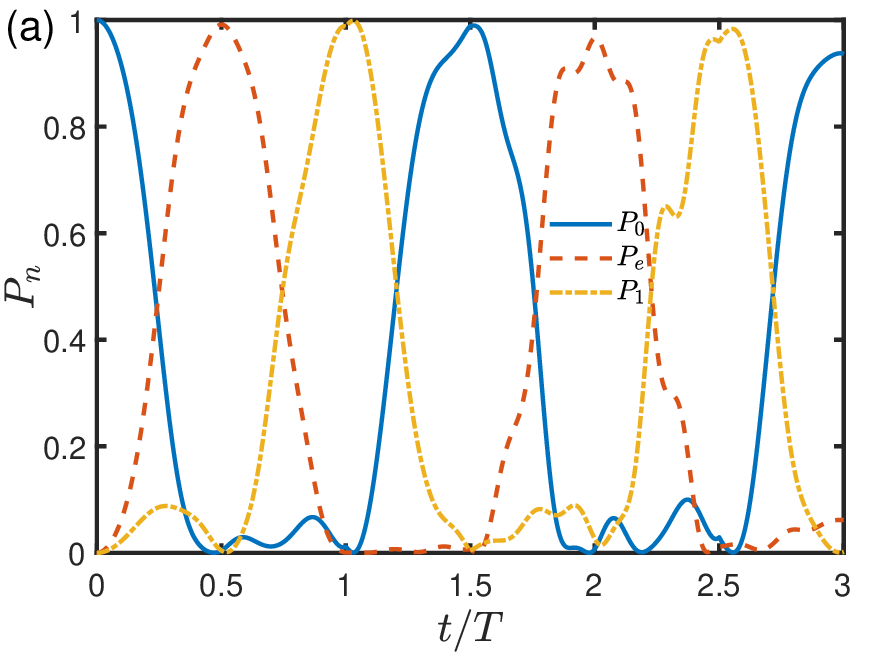}
\includegraphics[width=0.85\linewidth]{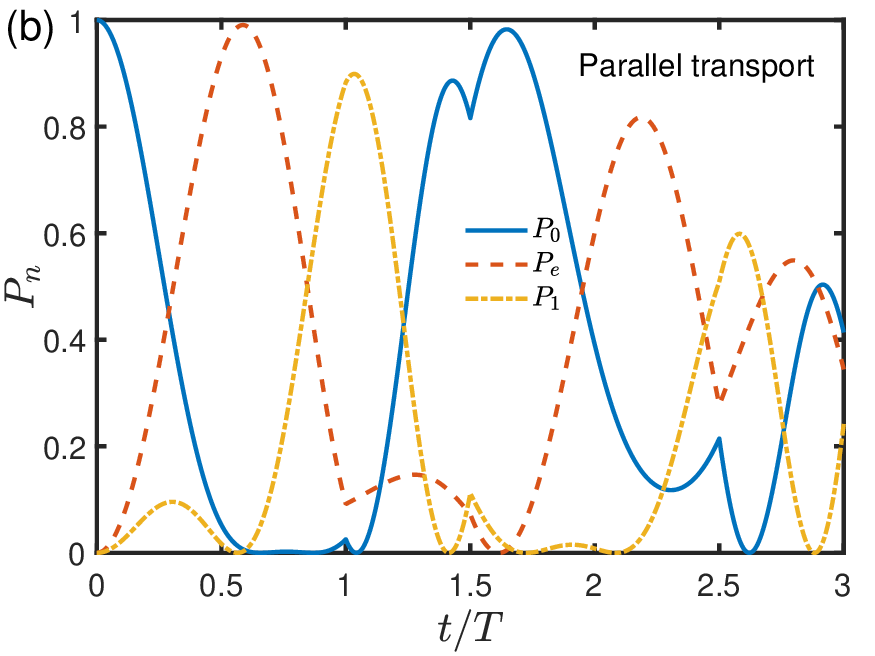}
\caption{State populations $P_n(t)$, with $n=0,1,e$, under the local error in Eq.~(\ref{delta1}) with $\epsilon=-0.2$, versus the evolution time $t/T$ during the cyclic population transfer. The global phase $f(t)$ satisfies the condition in Eq.~(\ref{Simplef}) with (a) $\lambda=5$ and (b) $\lambda=0$. The other parameters are the same as those in Fig.~\ref{CycRabi}.}\label{CycDetu}
\end{figure}

We compare the population dynamics $P_n$ in the presence of the global error in Eq.~(\ref{globalerror}) with $\epsilon=-0.2$ controlled by our protocol [see Fig.~\ref{CycRabi}(a)] and the conventional parallel-transport condition [see Fig.~\ref{CycRabi}(b)]. The former clearly outperforms the latter. During the first loop in Fig.~\ref{CycRabi}(a) with a moderate $\lambda=5$, the initial population in the state $|0\rangle$ can be transferred to state $|e\rangle$ with $P_e(T/2)=0.994$. Then it can be further transferred to $|1\rangle$ with $P_1(T)=0.994$ and back to $|0\rangle$ with $P_0(3T/2)=0.987$. During the second loop, it is found that $P_e(2T)=0.990$, $P_1(5T/2)=0.991$, and $P_0(3T)=0.989$. This implies that the nearly perfect loops can be infinitely continued under the dynamical correction. In sharp contrast, the cyclic population transfer under the parallel transport condition, i.e., $\lambda=0$, is subject to severe impact caused by the systematic error as shown in Fig.~\ref{CycRabi}(b). We find $P_e(T/2)=0.898$, $P_1(T)=0.673$, and $P_0(3T/2)=0.750$. Although $P_e(0.62T)=0.984$ and $P_0(1.70T)=0.990$ can be obtained as peak values, the cyclic population transfer can reluctantly last at most one loop with a deviating period.

In the presence of local error in Eq.~(\ref{delta1}) with $\epsilon=-0.2$, the dynamics of the state populations $P_n$'s are demonstrated with our protocol and that with parallel transport in Fig.~\ref{CycDetu}. We see that the adverse effects from local errors can be significantly corrected by our protocols in both populations and periods in comparison to those under the parallel-transport condition. As shown in Fig.~\ref{CycDetu}(a) with $\lambda=5$, the population are $P_e(T/2)=0.991$, $P_1(T)=0.990$, and $P_0(3T/2)=0.990$ during the first loop. We still have $P_e(2T)=0.966$, $P_1(5T/2)=0.960$, and $P_0(3T)=0.937$ during the second loop. In Fig.~\ref{CycDetu}(b) with $\lambda=0$, we see that $P_e(T/2)=0.931$, $P_1(T)=0.890$, and $P_0(3T/2)=0.816$. However, the deviating peak values can be found as $P_e(0.60T)=0.990$ and $P_0(1.65T)=0.982$. The results of the second loop cannot be regarded as a genuine population transfer.

\section{Discussion and Conclusion}\label{conclusion}

In summary, we propose a dynamical-correction protocol under our universal framework for controlling the time-dependent discrete-variable quantum systems within an ancillary picture, that can address arbitrary systematic errors in the control parameters. In the ideal situation, nonadiabatic paths can be generated by the von Neumann equation for the ancillary projection operators. In the presence of systematic errors, either global or local, undesirable transitions among those paths can be suppressed by the time-modulated global phase that, to a certain extent, has been overlooked by existing protocols, such as shortcuts to adiabaticity and holonomic quantum computation. In application, we demonstrate the unidirectional and cyclic population transfer in the ubiquitous three-level system using our dynamical correction. Our work offers a reliable error-correction tool in the noisy intermediate-scale quantum era and enhances the robustness of quantum network based on discrete-variable systems.

From the ansatz for the three-level system in Eq.~(\ref{Anci}), one can infer that two more independent control parameters (one for the local or global phase and another for the population) can be presented when adding one more level to the interested time-dependent system. For example, Eq.~(\ref{Anci}) could be straightforwardly extended to the generic two-band systems~\cite{Jin2024Entangling} of $M+N$ levels, allowing one to obtain at least one universal path that is robust against the deviations in the control parameters.

Under special parametric settings, our protocol can be transformed to various control protocols, such as stimulated Raman adiabatic passage, shortcut to adiabaticity, and holonomic transformation (which, based on the parallel transport actually assumes a constant global phase), since it has inherited the exceptional versatility of our universal control framework~\cite{Jin2024Shortcut}. Consequently, it is applicable to all platforms that can be controlled in the time domain, such as the superconducting system~\cite{Antti2019Superadiabatic}, ultracold atomic systems~\cite{Du2016Experimental}, and nitrogen-vacancy centers in diamond~\cite{Zhou2017Accelerated}. For instance, the model in Fig.~\ref{model} could be realized by a superconducting transmon qutrit~\cite{Antti2019Superadiabatic}. The detunings, the Rabi frequencies, and the phases of the driving fields can be individually controlled by the microwave pulses sent to a gate line that is capacitively coupled to the transmon.

\section*{Acknowledgment}

We acknowledge grant support from the Science and Technology Program of Zhejiang Province (No. 2025C01028).

\appendix

\section{Derivation details of the perturbed time-evolution operator}\label{DetailEvo}

This Appendix provides the details of the derivation of the time-evolution operator $U_I(t)$ in Eq.~(\ref{Uapprox}) in the presence of systematic errors. Using the Magnus expansion~\cite{Blanes2009Magnus,Liu2021Super} for the general time-evolution operator in Eq.~(\ref{Uformu}), we have
\begin{equation}\label{Umagnus1}
U_I(t)=\exp\left[\Lambda(t)\right], \quad \Lambda(t)=\sum_{l=1}^{\infty}\Lambda_l(t),
\end{equation}
where the first- and second-order terms are
\begin{equation}
\begin{aligned}
\Lambda_1(t)&=-i\epsilon\int_0^tdt_1\tilde{\mathcal{D}}(t_1),\\
\Lambda_2(t)&=\frac{(-i\epsilon)^2}{2}\int_0^tdt_1\int_0^{t_1}dt_2\left[\tilde{\mathcal{D}}(t_1), \tilde{\mathcal{D}}(t_2)\right],
\end{aligned}
\end{equation}
respectively. The first-order and second-order derivatives of $U_I(t)$ with respect to the perturbation coefficient $\epsilon$ are given by~\cite{Blanes2009Magnus}
\begin{equation}\label{UmagnusComm1}
\begin{aligned}
\frac{\partial U_I(t)}{\partial\epsilon}&=U_I(t)\sum_{k=0}^{\infty}\frac{1}{(k+1)!}ad_{\Lambda(t)}^k\left[\frac{\partial\Lambda(t)}{\partial\epsilon}\right],\\
\frac{\partial^2 U_I(t)}{\partial\epsilon^2}&=\frac{\partial U_I(t)}{\partial\epsilon}\sum_{k=0}^{\infty}\frac{1}{(k+1)!}ad_{\Lambda(t)}^k\left[\frac{\partial\Lambda(t)}{\partial\epsilon}\right]\\
&+U_I(t)\frac{\partial}{\partial\epsilon}\left\{\sum_{k=0}^{\infty}\frac{1}{(k+1)!}ad_{\Lambda(t)}^k\left[\frac{\partial\Lambda(t)}{\partial\epsilon}\right]\right\},
\end{aligned}
\end{equation}
respectively, where $ad_B^0(A)\equiv A$ and $ad_B^{k+1}(A)=[B, ad_B^{k}(A)]$. Using Eq.~(\ref{UmagnusComm1}), we have
\begin{widetext}
\begin{equation}\label{DetailDer}
\begin{aligned}
&U_I(t)=1+\left[\frac{\partial U_I(t)}{\partial \epsilon}\right]_{\epsilon=0}\epsilon+\left[\frac{\partial^2U_I(t)}{\partial\epsilon^2}\right]_{\epsilon=0}
\frac{\epsilon^2}{2}+\cdots=1+\left\{U_I(t)\sum_{k=0}^{\infty}\frac{1}{(k+1)!}ad_{\Lambda(t)}^k
\left[\frac{\partial\Lambda(t)}{\partial\epsilon}\right]\right\}_{\epsilon=0}\epsilon\\
&+\Big\{\frac{\partial U_I(t)}{\partial\epsilon}\sum_{k=0}^{\infty}\frac{1}{(k+1)!}ad_{\Lambda(t)}^k
\left[\frac{\partial\Lambda(t)}{\partial\epsilon}\right]+U_I(t)\frac{\partial}{\partial\epsilon}\left\{\sum_{k=0}^{\infty}\frac{1}{(k+1)!}
ad_{\Lambda(t)}^k\left[\frac{\partial\Lambda(t)}{\partial\epsilon}\right]\right\}\Big\}_{\epsilon=0}
\frac{\epsilon^2}{2}+\cdots\\
&\approx1+\left\{U_I(t)\left[\frac{\partial\Lambda(t)}{\partial\epsilon}+\frac{1}{2}
\left[\Lambda(t),\frac{\partial\Lambda(t)}{\partial\epsilon}\right]+\cdots\right]\right\}_{\epsilon=0}\epsilon
+\Big\{\frac{\partial U_I(t)}{\partial\epsilon}\left[\frac{\partial\Lambda(t)}{\partial\epsilon}+\frac{1}{2}
\left[\Lambda(t),\frac{\partial\Lambda(t)}{\partial\epsilon}\right]+\cdots\right]\\
&+U_I(t)\left[\frac{\partial^2\Lambda(t)}{\partial\epsilon^2}
+\frac{1}{2}\left[\frac{\partial\Lambda(t)}{\partial\epsilon},\frac{\partial\Lambda(t)}{\partial\epsilon}\right]
+\frac{1}{2}\left[\Lambda(t),\frac{\partial^2\Lambda(t)}{\partial\epsilon^2}\right]+\cdots\right]\Big\}_{\epsilon=0}
\frac{\epsilon^2}{2}\\
&=1-i\epsilon\mathcal{M}(t)-\frac{\epsilon^2}{2}
\left\{\mathcal{M}(t)^2+\int_0^tdt_1\int_0^{t_1}dt_2\left[\tilde{\mathcal{D}}(t_1), \tilde{\mathcal{D}}(t_2)\right]\right\},
\end{aligned}
\end{equation}
\end{widetext}
where $\mathcal{M}(t)$ is the error rotation defined in Eq.~(\ref{Convension}) and the approximation holds for $\epsilon\ll1$. Several identities were used in the above derivation:
\begin{equation}
  \begin{aligned}
    & U_I(t)|_{\epsilon=0}=1, \quad
    \frac{\partial U_I(t)}{\partial\epsilon}|_{\epsilon=0}=-i\mathcal{M}(t), \\
    & \Lambda(t)|_{\epsilon=0}=0, \quad
    \frac{\partial\Lambda(t)}{\partial\epsilon}|_{\epsilon=0}=-i\mathcal{M}(t), \\
    & \frac{\partial^2\Lambda(t)}{\partial\epsilon^2}|_{\epsilon=0}
    =-\int_0^tdt_1\int_0^{t_1}dt_2[\tilde{\mathcal{D}}(t_1), \tilde{\mathcal{D}}(t_2)].
  \end{aligned}
\end{equation}

\section{Derivation details of fidelity in the presence of systematic errors}\label{DetailOverlap}

This Appendix provides a detailed derivation of the state fidelity in Eq.~(\ref{overlap}). With $U_I(t)$ in Eq.~(\ref{Uapprox}) or (\ref{DetailDer}), the fidelity or overlap between the instantaneous state governed by $U_I(t)$ and the target state can be expressed as
\begin{widetext}
\begin{equation}\label{overlap1}
\begin{aligned}
&\mathcal{F}\equiv\left|\langle\mu_k(0)|U_I(t)|\mu_k(0)\rangle\right|^2\approx\Big|1-i\epsilon\mathcal{M}_{kk}(t)-\frac{\epsilon^2}{2}\sum_{n=1}^K
\Big\{\langle\mu_k(0)|\mathcal{M}(t)|\mu_n(0)\rangle\langle\mu_n(0)|\mathcal{M}(t)|\mu_k(0)\rangle\\
+&\int_0^tdt_1\int_0^{t_1}dt_2\langle\mu_k(0)|\left[\tilde{\mathcal{D}}(t_1)|\mu_n(0)\rangle\langle\mu_n(0)|
\tilde{\mathcal{D}}(t_2)-\tilde{\mathcal{D}}(t_2)|\mu_n(0)\rangle\langle\mu_n(0)|\tilde{\mathcal{D}}(t_1)\right]
|\mu_k(0)\rangle\Big\}\Big|^2\\
=&\left|1-i\epsilon\mathcal{M}_{kk}(t)-\frac{\epsilon^2}{2}\sum_{n=1}^K\left\{\mathcal{M}_{kn}(t)\mathcal{M}_{nk}(t)
+\int_0^tdt_1\int_0^{t_1}dt_2\left[\tilde{\mathcal{D}}_{kn}^{\rm err}(t_1)\tilde{\mathcal{D}}_{nk}^{\rm err}(t_2)-\tilde{\mathcal{D}}_{kn}^{\rm err}(t_2)\tilde{\mathcal{D}}_{nk}^{\rm err}(t_1)\right]\right\}\right|^2\\
=&\left|1-i\epsilon\mathcal{M}_{kk}(t)-\epsilon^2\sum_{n=1}^K\int_0^tdt_1\mathcal{M}_{nk}(t_1)
\tilde{\mathcal{D}}_{kn}^{\rm err}(t_1)\right|^2\\
=&1+\epsilon^2\mathcal{M}_{kk}^2(t)-\epsilon^2\sum_{n=1}^K\int_0^tdt_1
\left[\mathcal{M}_{nk}(t_1)\tilde{\mathcal{D}}_{kn}^{\rm err}(t_1)
+\mathcal{M}_{kn}(t_1)\tilde{\mathcal{D}}_{nk}^{\rm err}(t_1)\right]+\mathcal{O}\left(\epsilon^3\right)\\
=&1-\epsilon^2\sum_{n=1,n\ne k}^K\left|\mathcal{M}_{kn}(t)\right|^2+\mathcal{O}\left(\epsilon^3\right),
\end{aligned}
\end{equation}
\end{widetext}
where $\mathcal{M}_{kn}(t)\equiv\langle\mu_k(0)|\mathcal{M}(t)|\mu_n(0)$ and $\tilde{\mathcal{D}}_{kn}^{\rm err}(t)\equiv\langle\mu_k(0)|\tilde{\mathcal{D}}(t)|\mu_n(0)\rangle$. The step from the fourth line to the fifth line of Eq.~(\ref{overlap1}) employs the relations
\begin{equation}
\begin{aligned}
&\int_0^tdt_1\int_0^{t_1}dt_2\tilde{\mathcal{D}}_{kn}^{\rm err}(t_1)\tilde{\mathcal{D}}_{nk}^{\rm err}(t_2)\\
&=\int_0^tdt_1\tilde{\mathcal{D}}_{kn}^{\rm err}(t_1)\left[\int_0^{t_1}dt_2\tilde{\mathcal{D}}_{nk}^{\rm err}(t_2)\right]\\
&=\int_0^tdt_1\mathcal{M}_{nk}(t_1)\tilde{\mathcal{D}}_{kn}^{\rm err}(t_1),
\end{aligned}
\end{equation}
and
\begin{equation}
\begin{aligned}
&\int_0^tdt_1\int_0^{t_1}dt_2\tilde{\mathcal{D}}_{kn}^{\rm err}(t_2)\tilde{\mathcal{D}}_{nk}^{\rm err}(t_1)\\ &=\int_0^t\mathcal{M}_{kn}(t_1)d\left[\mathcal{M}_{nk}(t_1)\right]\\
&=\mathcal{M}_{kn}(t)\mathcal{M}_{nk}(t)
-\int_0^tdt_1\mathcal{M}_{nk}(t_1)\tilde{\mathcal{D}}_{kn}^{\rm err}(t_1).
\end{aligned}
\end{equation}

\bibliographystyle{apsrevlong}
\bibliography{ref}

\end{document}